\documentclass[journal]{IEEEtran} 
\newcommand{\imgwidth}{0.7} 

\usepackage{graphicx}
\usepackage{amsmath}
\usepackage{amssymb}
\usepackage{setspace}
\usepackage{cite}
\usepackage[printonlyused]{acronym}

\usepackage{lastpage} 
\usepackage{eso-pic}  

\usepackage{lastpage} 
\usepackage{eso-pic}  
\usepackage{hyperref}
\AddToShipoutPicture{
\put(0,15){ 
\begin{minipage}{\paperwidth}
\footnotesize
\hspace{1in}\hspace{\oddsidemargin}
\hfill
\thepage\,/\,\pageref*{LastPage}
\hspace{1in}\hspace{\oddsidemargin}
\vspace{10ex}
\end{minipage}
}}

\acrodef{DoA}{Direction of Arrival}
\acrodef{MUSIC}{MUltiple SIgnal Classification}
\acrodef{STFT}{Short-Time Fourier Transform}
\acrodef{SNR}{Signal-to-Noise Ratio}
\acrodef{ASR}{Automatic Speech Recognition}
\acrodef{SH}{Spherical Harmonic}
\acrodef{PWD}{Plane-Wave Density}
\acrodef{SH-MUSIC}{Spherical Harmonics MUltiple SIgnal Classification}
\acrodef{LTI}{Linear Time-Invariant}
\acrodef{HRIR}{Head-Related Impulse Response}
\acrodef{HRTF}{Head-Related Transfer Function}
\acrodef{STD}{STandard Deviation}
\acrodef{BEM}{Boundary Element Method}
\acrodef{SFT}{Spherical Fourier Transform}
\acrodef{MTF}{Multiplicative Transfer Function}
\acrodef{DPD}{Direct Path Dominance}
\acrodef{AM}{Amplitude Modulation}
\acrodef{FM}{Frequency Modulation}
\newcommand{\bb}[1]{\mathbf{#1}}

\title{Direction of arrival estimation using microphone array processing for moving humanoid robots}
\author{V.~Tourbabin*,~\IEEEmembership{Student~Member,~IEEE} and B.~Rafaely,~\IEEEmembership{Senior~Member,~IEEE} 
\date{}
\thanks{
The authors are with the Department of Electrical and Computer Engineering, Ben-Gurion University of the Negev, Be'er-Sheva 84105, Israel (email: \{tourbabv,br\}@ee.bgu.ac.il)}}

\begin{document}
	\maketitle
	\begin{abstract} 
The auditory system of humanoid robots has gained increased attention in recent years. 
This system typically acquires the surrounding sound field by means of a microphone array.
Signals acquired by the array are then processed using various methods.
One of the widely applied methods is direction of arrival estimation.
The conventional direction of arrival estimation methods assume that the array is fixed at a given position during the estimation. However, this is not necessarily true for an array installed on a moving humanoid robot. The array motion, if not accounted for appropriately, can introduce a significant error in the estimated direction of arrival.
The current paper presents a signal model that takes the motion into account. Based on this model, two processing methods are proposed. The first one compensates for the motion of the robot. The second method is applicable to periodic signals and utilizes the motion in order to enhance the performance to a level beyond that of a stationary array.
Numerical simulations and an experimental study are provided, demonstrating that the motion compensation method almost eliminates the motion-related error. It is also demonstrated that by using the motion-based enhancement method it is possible to  improve the direction of arrival estimation performance, as compared to that obtained when using a stationary array.
\end{abstract}



\section{Introduction} 
\IEEEPARstart{A}{udition} is an essential part of humanoid robots; it facilitates the robot's communication with the environment by exploiting the surrounding sound field.
The auditory system of a humanoid robot is typically comprised of a microphone array and a set of signal processing methods. 
One of the widely applied array processing methods is \ac{DoA} estimation. This method is used for sound source localization\cite{Pavlaidi2013}, spatial filtering and dereverberation \cite{Peled2013}, as well as for preprocessing in \ac{ASR} systems \cite{Nakamura2013}.
Several different approaches for \ac{DoA} estimation are used in the literature, including time-delay-based algorithms 
\cite{Alameda2014},  beamforming \cite{Gershman1995}, maximum likelihood estimators \cite{Zhang2008} and subspace-based algorithms 
\cite{Schmidt1986,Roy1989}. 
These algorithms usually assume that the microphone array is fixed at a given position. 
However, a microphone array installed on a robot is not necessarily fixed; it can move in accordance with the robot's activity. 
Motion of the robot poses several challenges for \ac{DoA} estimation algorithms. One of the problems, associated with motion and addressed in the literature, is the ego-motion noise originating from robot motors and joint movements \cite{Ince2011}. 
Another problem, addressed in the current paper, is related to the motion of the array relative to the sound field. For a moving array, the \ac{DoA}s to be estimated are continuously changing during the data acquisition. Thus, direct application of existing \ac{DoA} estimation methods, based on stationary array models, may lead to severe degradation in performance.

One possible approach that can overcome this problem is the ``\emph{stop-perceive-act}'' principle \cite{Nakadai2000}, which suggests stopping the robot while acquiring data for the estimation. This approach can also reduce the ego-motion noise. However, it may prevent the robot from receiving new commands while moving as a response to a previous command, thereby imposing behavioral constraints that may limit natural robot interaction with the environment.
An alternative approach is to take the motion into account by utilizing the information about the array position, speed and acceleration, as a function of time. This information can be obtained from the operating system of the robot.
Incorporating this information into the processing model can reduce the \ac{DoA} estimation errors originating from array motion. Moreover, the motion can be utilized to improve the performance beyond that of a stationary array. For example, it is possible to reduce spatial aliasing in beamforming by using a linear array moving with constant acceleration \cite{Chang2011} or by using a planar array rotating with constant angular velocity \cite{Cigata2008}. It is also possible to improve the \ac{DoA} estimation resolution by using the synthetic aperture technique, which has been applied to underwater linear arrays towed at a constant speed (see, for example, \cite{Yen1989}). In addition, when the sampling of a field is considered, the motion of the sensors can be utilized to reduce the reconstruction error \cite{Unnikrishnan2013}.
The above mentioned techniques provide an insight into a possible approach for the processing of moving microphone arrays for robot audition. However, as mentioned above, these techniques are usually limited to linear arrays moving along a straight line or planar arrays rotating with constant velocity. Therefore, these techniques cannot be directly applied to \ac{DoA} estimation using arbitrarily moving arrays of general geometry, which is usually the case in the context of humanoid robot audition. This problem is addressed in the current paper.

The approach adopted here exploits the sphere-like shape of the humanoid robot head, which facilitates processing in the \ac{SH} domain. 
It is shown that processing in the \ac{SH} domain, in contrast to the more traditional space-domain processing, enables a convenient representation of robot's motion by a sequence of relatively simple linear transformations. 
This linear representation is utilized here to reduce the motion-induced error in \ac{DoA} estimation.
In addition, an approach is presented for periodic signals that improves the \ac{DoA} estimation beyond that of a stationary array. The improvement is achieved by combining the measurements from different spatial positions in a way similar to the synthetic aperture technique \cite{Tourbabin2014}. 
It is emphasized that in the SH domain, the motion-aware processing is applied to the \ac{PWD} function of the surrounding sound field, which is obtained prior to the \ac{DoA} estimation. Hence, the proposed methods are generic, i.e., they can be incorporated into various \ac{DoA} estimation techniques.

The remainder of the paper is organized as follows. Section \ref{sec:background} introduces the stationary array model and the \ac{DoA} estimation algorithm to be used for the evaluation of the proposed approach. Then, in sections \ref{sec:moving} and \ref{sec:methods}, the proposed approach is presented and analyzed. Sections \ref{sec:numercompensation} and \ref{sec:numerenhancement} provide numerical investigations of the two proposed methods. Section \ref{sec:experiment} validates the proposed methods with experimental data using a humanoid robot. Finally, section \ref{sec:conclusion} concludes the paper.

\section{Background} 
\label{sec:background}
This section starts with the derivation of a signal model for a stationary microphone array. A \ac{DoA} estimation algorithm is also outlined in this section, for completeness. This algorithm will be used for assessing the performance of the proposed methods. Later, in section \ref{sec:moving}, the signal model derived here will be extended to account for array motion.
\subsection{Stationary array model} 
Consider an array of $M$ microphones distributed on the surface of a robot head. The array center coincides with the origin of a standard spherical coordinate system \cite{Arfken2005} denoted by $(r,\theta,\phi)$, representing the radius, elevation and azimuth, respectively. For the moment, suppose that the surrounding sound field is produced by a single far-field source located in the direction $(\theta_s,\phi_s)$. In the absence of the array, the sound pressure produced by this source at the origin of the coordinate system is denoted by $a(t,\theta_s,\phi_s)$. Using this notation, the time-sampled pressure at all microphones $\bb{p}(t)=[p_1(t)\,p_2(t)\,\cdots\,p_M(t)]^T$ can be expressed as
\begin{equation}
	\bb{p}(t)=\sum\limits_{\tau=-\infty}^{\infty}{\bb{h}(t-\tau,\theta_s,\phi_s)a(\tau,\theta_s,\phi_s)}+\bb{n}(t),
	\label{eq:b1}
\end{equation}
where $(\cdot)^T$ denotes the transpose operator and $\bb{n}(t)$ represents additive noise. Vector $\bb{h}(t,\theta_s,\phi_s)=[h_1(t,\theta_s,\phi_s)\,h_2(t,\theta_s,\phi_s)\,\cdots\,h_M(t,\theta_s,\phi_s)]^T$ contains the impulse responses of the appropriate  discrete \ac{LTI} systems. These systems describe the transformation of the sound field from the origin of the coordinate system to the microphones. In the context of the human auditory system, the impulse response is usually referred to as the \ac{HRIR} \cite{Blauert1997}.  In order to generalize the expression in \eqref{eq:b1}, suppose that instead of a single source, the sound field is produced by an arbitrary distribution of far-field sources, with their relative contribution to the sound field at the origin given by $a(t,\theta,\phi)$. In this case, the output of the microphones can be obtained by integrating over all directions, i.e.,
\begin{align}
	\nonumber\bb{p}(t)&=\int\limits_{0}^{2\pi}\int\limits_{0}^{\pi}{\sum\limits_{\tau=-\infty}^{\infty}{\bb{h}(t-\tau,\theta,\phi)
	a(\tau,\theta,\phi)}\sin\theta d\theta d\phi}\\
	&+ \bb{n}(t).
	\label{eq:b2}
\end{align}
\ac{DoA} estimation is usually performed using the \ac{STFT} of the microphone outputs
\begin{equation}
	\bb{p}(i,\omega)=\sum\limits_{t=0}^{T-1}{w(t)\bb{p}(t+iD)e^{-j\frac{2\pi}{T}\omega t}},
	\label{eq:b3}
\end{equation}
where $T$ is the duration of the transform frame in samples, $D$ is the offset between subsequent frames, $w(t)$ is a window of duration $T$, $j=\sqrt{-1}$, $i$ is the index of the time frame, and $\omega=0,1,...,T-1$ is the frequency bin index. By applying the \ac{STFT} on both sides of \eqref{eq:b2} and using the \ac{MTF} approximation \cite{Avargel2007}, we obtain
\begin{equation}
	\bb{p}(i,\omega)=\int\limits_{0}^{2\pi}\int\limits_{0}^{\pi}{\bb{v}^*(\omega,\theta,\phi)a(i,\omega,\theta,\phi)
	\sin\theta d\theta d\phi}+\bb{n}(i,\omega),
	\label{eq:b4}
\end{equation}  
where $a(i,\omega,\theta,\phi)$ is the \ac{STFT} of $a(t,\theta,\phi)$, being the \ac{PWD} function of the sound field in the time frame $i$. 
Vector $\bb{v}^*(\omega,\theta,\phi)$ is the Fourier transform of $\bb{h}(t,\theta,\phi)$, holding the direction-dependent responses of all the microphones at a given frequency, and is known as the array steering vector in the array processing literature \cite{VanTrees2002}. Finally, $\bb{n}(i,\omega)$ is the \ac{STFT} of $\bb{n}(t)$ and $(\cdot)^*$ denotes the complex-conjugate operator. Recall that in \eqref{eq:b4} we use the \ac{MTF} approximation, which implies that time-domain convolution between the source signal in each time frame and the impulse response of a system can be approximated as their multiplication in the \ac{STFT} domain.  This imposes a constraint on the duration of the time frames $T$, which should be sufficiently large compared to the duration of the \ac{HRIR}, $\bb{h}(t)$ \cite{Avargel2007}. The duration of the human \ac{HRIR} is in the order of milliseconds. Assuming a similar duration for the humanoid-robot \ac{HRIR}, it is believed that a time frame longer than $10$ ms should result in a reasonable approximation. 

The integral in \eqref{eq:b4} can be rewritten as a sum in the \ac{SH} domain using Parseval's theorem for the \ac{SFT}\cite{Arfken2005}:
\begin{equation}
	\bb{p}(i,\omega)=\sum\limits_{n=0}^{N}{\sum\limits_{m=-n}^{n}{\bb{v}_{nm}^*(\omega)a_{nm}(i,\omega)}}+\bb{n}(i,\omega),
	\label{eq:b5}
\end{equation}
where $a_{nm}(i,\omega)$ and $\bb{v}_{nm}(\omega)$ are the \ac{SFT} coefficients of the \ac{PWD} function and of the complex conjugate of the steering vector, respectively. Note that in \eqref{eq:b5} it is assumed that the \ac{SH} order of the sound field on the robot head surface is limited to $N$. The maximum \ac{SH} order contained in the field, $N$, is a function of frequency and depends on the geometry of the head surface. It may be difficult to derive the expression for $N$ considering a general head geometry. Nevertheless, assuming that the head surface is close to spherical, it is suggested here to use the expression for the effective order of the pressure on the surface of a rigid sphere, $N=\lceil kr\rceil$\cite{Rafaely2005}, where $k=2\pi f_s \omega/Tc$ is the wavenumber, $\lceil\cdot\rceil$ is the ceiling operator, $c$ denotes the speed of sound, and $f_s$ is the sampling frequency in Hz. The dependence of $N$ on frequency is omitted for notation simplicity.

Note that \eqref{eq:b5} implies that the measured sound field is effectively represented by the first $N$ orders of the \ac{PWD} function. This allows us to rewrite \eqref{eq:b5} in a matrix form:
\begin{equation}
	\bb{p}(i,\omega)=\bb{V}(\omega)\bb{a}(i,\omega)+\bb{n}(i,\omega),
	\label{eq:b6}
\end{equation}
where $\bb{V}(\omega)=[\bb{v}_{0,0}^*(\omega)\,\bb{v}_{1,-1}^*(\omega)\,\bb{v}_{1,0}^*(\omega)\,\cdots\,\bb{v}_{N,N}^*(\omega)]\in\mathbb{C}^{M\times (N+1)^2}$ and\\ $\bb{a}(i,\omega)=[a_{0,0}(i,\omega)\,a_{1,-1}(i,\omega)\,a_{1,0}(i,\omega)\,\cdots\,a_{N,N}(i,\omega)]^T\in\mathbb{C}^{(N+1)^2\times 1}$.
The model in \eqref{eq:b6} relates the sound field, represented by its \ac{PWD} function in the \ac{SH} domain $\bb{a}(i,\omega)$, to the array measurements $\bb{p}(i,\omega)$. Processing of the array based on this model requires knowledge of the array steering matrix $\bb{V}(\omega)$. In practice, this matrix can be obtained from measurements \cite{Maazaoui2012} or by numerical simulation \cite{Tourbabin2014Dec}. The model in \eqref{eq:b6} provides the basis for the discussion in the following sections.

\subsection{SH-MUSIC} 
\label{sec:music}
This section outlines the \ac{SH-MUSIC} \cite{Khaykin2009} algorithm for \ac{DoA} estimation in the \ac{SH} domain. It is provided here for convenience as a reference algorithm for the evaluation of the methods proposed below.

Assume that a sound field produced by $S$ spatially separated far-field sources is sampled using an array of $M>S$ microphones. The \ac{SH-MUSIC} algorithm for the estimation of the \ac{DoA}s of all the sources proceeds as follows:\\
\emph{Step $1$:}
Obtain $J\geq S$ snapshots of the sound pressure at all microphones $\bb{p}(i,\omega),\,i=1,2,...,J$, by calculating the \ac{STFT} of the microphone outputs, as defined in \eqref{eq:b3}. 
\\
\emph{Step $2$:}
Estimate the \ac{PWD} functions $\bb{a}(i,\omega)$. 
Considering the model in \eqref{eq:b6}, this step can be accomplished by
\begin{equation}
	\hat{\bb{a}}(i,\omega)=\bb{V}^{\dag}(\omega)\bb{p}(i,\omega),
	\label{eq:b7}
\end{equation} 
where $(\cdot)^{\dag}$ denotes the Moore-Penrose pseudo-inverse \cite{Golub1996}. 
Recall that the sound field on the surface of the robot head is assumed to have a limited \ac{SH} order $N$, implying that $\bb{V}(\omega)\in\mathbb{C}^{M\times (N+1)^2}$. The maximum \ac{SH} order depends on frequency and, as suggested above, is given by $N=\lceil kr \rceil$. Hence, in the frequency range for which $M>(N+1)^2$, equation \eqref{eq:b7} will result in the Least-Squares sense estimation, while at higher frequencies, the estimation will be in the Minimum-Norm sense \cite{Golub1996}.
\\
\emph{Step $3$:}
Estimate the modal narrow-band covariance matrix at each frequency by averaging over time:
\begin{equation}
	\bb{Q}(\omega)=\frac{1}{J}\sum\limits_{i=1}^J{\hat{\bb{a}}(i,\omega)\hat{\bb{a}}^H(i,
	\omega)},
	\label{eq:b8}
\end{equation}
where $(\cdot)^H$ denotes the conjugate-transpose operator. Next, average the narrow-band covariance matrices over frequency:
\begin{equation}
	\tilde{\bb{Q}}=\frac{1}{T}\sum\limits_{\omega=1}^T{\bb{Q}(\omega)}.
	\label{eq:b9}
\end{equation}
Note that the averaging in \eqref{eq:b9} can be performed over a selected subset of frequencies. The averaging over frequency (also known as frequency smoothing) preserves the spatial structure of the covariance matrix, due to the decoupling between space and frequency dependent parameters that is inherent to the \ac{SH} domain \cite{Khaykin2009}.
\\
\emph{Step 4:}
Estimate the noise subspace using $\tilde{\bb{Q}}$ and construct the MUSIC spectrum $P(\theta,\phi)$ \cite{Schmidt1986}. Note that this step may require whitening of the noise covariance matrix \cite{Khaykin2009}.
\\
\emph{Step 5:}
Obtain the \ac{DoA} estimates by picking $S$ values of $(\theta,\phi)$ corresponding to the $S$ greatest peaks of $P(\theta,\phi)$.

It is emphasized here that \ac{DoA} estimation in the \ac{SH} domain will usually start with the first two steps of the above described algorithm. The purpose of these steps is to estimate the \ac{SH} coefficients of the \ac{PWD} function to be used in steps $3-5$ for actual \ac{DoA} estimation.
The methods presented below aim to take array motion into account in the second step, thereby accounting for motion in any \ac{DoA} estimation algorithm that uses the \ac{PWD} function estimates, while \ac{SH-MUSIC} serves here as an example of an algorithm for evaluation purposes.

\section{Moving array model} 
\label{sec:moving}
The previous section presented an overview of a model for the processing of signals from stationary arrays. However, microphone arrays mounted on a moving humanoid robot are expected to move in accordance with the robot's activity. The current section is concerned with an extension of the model to account for the array motion, while the sources are assumed to be fixed at their positions. 

Assume that the origin of the coordinate system is positioned at the array center and moves together with the array. Thus, the sound field, as viewed from this coordinate system, is continuously moving in the direction reciprocal to the array motion. Similarly to in the stationary array case, the \ac{STFT} of the microphone outputs is calculated by dividing the time-line into frames of $T$ samples each. 
The effect of motion within each time frame \cite{Poletti2010} is neglected because the speed of the robot is believed to be very low relative to the speed of sound.
Nevertheless, the motion between the time frames is accounted for by adjusting the stationary array model:
\begin{equation}
	\bb{p}(i,\omega)=\bb{V}_i(\omega)\bb{W}_i(\omega)\bb{a}(i,\omega)+\bb{n}(i,\omega),
	\label{eq:m1}
\end{equation}
where $\bb{a}(i,\omega)$ are the \ac{SH} coefficients of the \ac{PWD} function measured by a stationary array located at a reference position and $\bb{W}_i(\omega)$ describes the transformation of the sound field due to the array motion between the reference position and its position during the $i^{th}$ time frame. 
Note that $\bb{W}_i(\omega)$ is not, in general, a square matrix, because the transformation may affect the \ac{SH} order of the field. Moreover, the order of the transformed field may vary between the time frames. Hence, in contrast to the stationary case, here, the number of columns in the steering matrix $\bb{V}_i(\omega)$ depends on $i$, as indicated by the subscript. Additional details on the \ac{SH} order of the transformed field are provided in section \ref{sec:translation}.

\begin{figure}[ht] 
	\centering
	\includegraphics[width=\columnwidth]{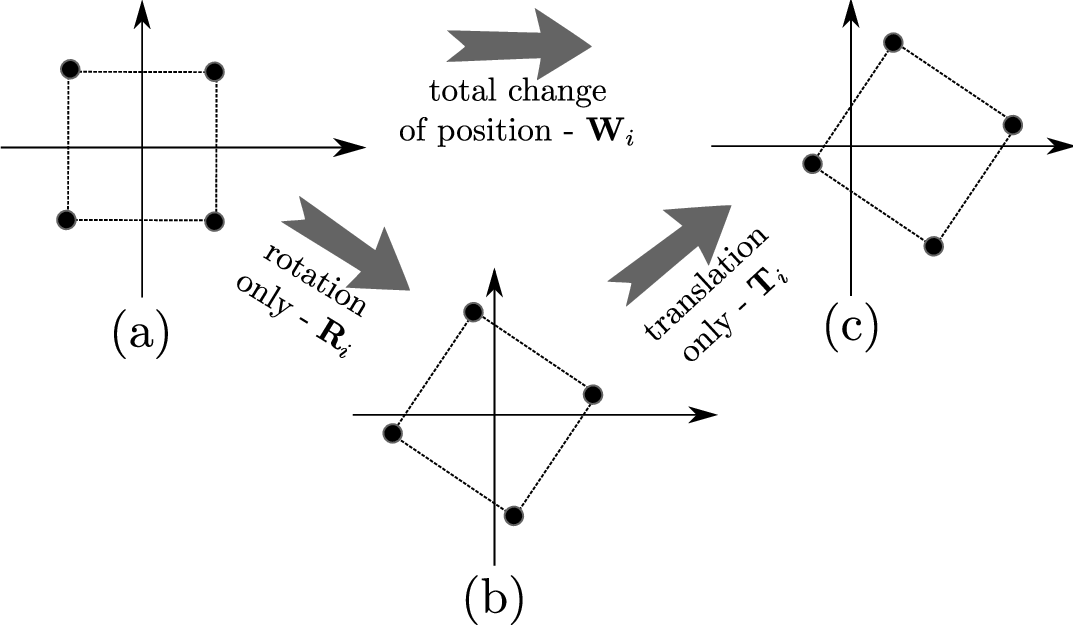}
	\caption{An illustration of a rectangular $4$-microphone array moving from the original position in (a) to the final position in 
	(c). The motion is divided into a rotation from (a) to (b) and a subsequent translation from (b) to (c).}
	\label{fig:m1}
\end{figure}

According to Chasles' theorem, the change in the array position induced by the motion of the robot between two time frames, can be divided into a rotation followed (or preceded) by a translation (see, for example, \cite{Heard2006}, page $42$). This implies that the transformation matrix $\bb{W}_i(\omega)$ can be decomposed as
\begin{equation}
	\bb{W}_i(\omega)=\bb{T}_i(\omega)\bb{R}_i(\omega),
	\label{eq:m2}
\end{equation}
where $\bb{R}_i(\omega)$ and $\bb{T}_i(\omega)$ describe the rotation and translation parts of the transformation, respectively.
This idea is illustrated schematically in Fig. \ref{fig:m1}.
In the \ac{SH} domain, the matrices $\bb{R}_i(\omega)$ and $\bb{T}_i(\omega)$ have closed form expressions as functions of the desired rotation angles and translation vector, respectively. These expressions are provided and discussed in the following two subsections.

\subsection{Array rotation} 
\label{sec:rotation}
The rotation matrix $\bb{R}_i$ is given by the Wigner-D matrix \cite{Kostelec2008}. 
This matrix is block-diagonal and unitary (see the appendix for the proof of unitarity) and is given by
\begin{equation}
	\bb{R}_i=\left(
	\begin{tabular}{llcl}
		$\bb{D}_{0}$ & $\bb{0}_{0,1}$ & $\cdots$ & $\bb{0}_{0,N}$ \\
		$\bb{0}_{1,0}$ & $\bb{D}_{1}$ & $\cdots$ & $\bb{0}_{1,N}$ \\
		$\vdots$ &  & $\ddots$ & $\vdots$ \\
		$\bb{0}_{N,0}$ & $\bb{0}_{N,1}$ & $\cdots$ & $\bb{D}_{N}$ 
		\end{tabular}
	\right),
	\label{eq:m3}
\end{equation}
where $\bb{0}_{m_1,m_2}$ is a zero matrix having $2m_1+1$ rows and $2m_2+1$ columns and $\bb{D}_n\in\mathbb{C}^{(2n+1)\times(2n+1)}$ is given by
\begin{equation}
	\bb{D}_n=\left(
	\begin{tabular}{ccc}
		$D_{-n,-n}^n$ & $\cdots$ & $D_{-n,n}^n$ \\
		$\vdots$ & $\ddots$ & $\vdots$ \\
		$D_{n,-n}^n$ & $\cdots$ & $D_{n,n}^n$ 
	\end{tabular}
	\right),
	\label{eq:m4}
\end{equation}
where $D_{m_1,m_2}^{m_3}$ is the short notation for the Wigner-D function \cite{Varshalovich1988} $D_{m_1,m_2}^{m_3}(\alpha_i,\beta_i,\gamma_i)$ with $\alpha_i$, $\beta_i$ and $\gamma_i$ being the Euler angles \cite{Arfken2005} of the rotation in the $i^{th}$ time frame. Note that $\bb{R}_i\in\mathbb{C}^{(N+1)^2\times(N+1)^2}$ is a square matrix, implying that rotation does not affect the \ac{SH} order of the field. Furthermore, observe that the rotation matrix does not depend on frequency, i.e., the transformation of the \ac{SH} coefficients due to a rotation by $(\alpha_i,\beta_i,\gamma_i)$ is identical for all frequency components. Hence, the dependence of $\bb{R}_i$ on $\omega$ will be omitted in the subsequent discussion. Further details on the Wigner-D function can be found in the appendix.

A special case of particular interest is when the head rotates in the horizontal plane. In this case, the rotation is completely specified by $\alpha_i$, while $\beta_i=\gamma_i=0$. Substituting into the expression for the Wigner-D function results in
\begin{align}
	\nonumber D_{m_1,m_2}^{m_3}(\alpha_i,0,0)&=e^{-jm_1\alpha_i}d_{m_1,m_2}^{m3}(0)e^{-jm_2 0}\\
	&=e^{-jm_1\alpha_i}\delta_{m_1,m_2},
	\label{eq:m5}
\end{align}  
where $d_{m_1,m_2}^{m3}(\cdot)$ and $\delta_{m_1,m_2}$ are the Wigner-d and Kronecker delta functions, respectively. The result in \eqref{eq:m5} implies that when the rotation is purely horizontal, the rotation matrix $\bb{R}_i$ is a diagonal matrix. Moreover, the diagonal values are obtained by a relatively simple evaluation of the exponential in \eqref{eq:m5}. In addition, note that for $\alpha_i\rightarrow 0$ the diagonal term $e^{-jm_1\alpha_i}\rightarrow 1$. Hence, as would be expected, $\bb{R}_i$ converges to the identity matrix $\bb{I}$.

\subsection{Array translation} 
\label{sec:translation}
The translation matrix $\bb{T}_i(\omega)$ describes the effect of the translation part of array motion between the reference and the $i^{th}$ time frame. Denote the \ac{SH} order and degree indices before the translation by $n,m$ and after the translation by $n',m'$. The element of $\bb{T}_i(\omega)$ in row $n'^2+n'+m'$ and column $n^2+n+m$ for far-field sources that are outside the measurement region, is given by \cite{Peleg2011}
\begin{equation}
	[\bb{T}_i]_{n'^2+n'+m'}^{n^2+n+m}=\sum\limits_{q=0}^{\lceil k r_i\rceil}{j_q(k r_i)\cdot Y_{q}^{m-m'}(\theta_i,\phi_i)
	\cdot C_{n',m'}^{n,m,q}},
	\label{eq:m6}
\end{equation}
where $j_q(\cdot)$ and $Y_q^{m-m'}(\cdot)$ are the spherical Bessel and spherical harmonic functions, respectively, $r_i$ and $(\theta_i,\phi_i)$ are the distance and the direction of the translation in the $i^{th}$ time window, and the dependence on $\omega$ is expressed through the wavenumber $k$, for convenience. Coefficient $C_{n',m'}^{n,m,q}$ is given by \cite{Peleg2011}
\begin{align}
	\nonumber C_{n',m'}^{n,m,q}&=4\pi j^{(n'+q-n)}(-1)^m\sqrt{\frac{(2n+1)(2n'+1)(2q+1)}{4\pi}}\\
	&\times 
	\left(
	\begin{tabular}{ccc}
		$n$ & $n'$ & $q$ \\
		$0$ & $0$ & $0$
	\end{tabular}
	\right)
	\left(
	\begin{tabular}{ccc}
		$n$ & $n'$ & $q$ \\
		$-m$ & $m'$ & $m-m'$
	\end{tabular}
	\right),
	\label{eq:m7}
\end{align}
where $\left(
	\begin{tabular}{ccc}
		$J_1$ & $J_2$ & $J_3$ \\
		$m_1$ & $m_2$ & $m_3$
	\end{tabular}
	\right)$
is the Wigner-$3$j symbol.
The sum in \eqref{eq:m6} is limited to $\lceil k r_i\rceil$ because $j_q(k r_i)\approx 0$ for $q>>k r_i$ \cite{Rafaely2005}. Note that the translation matrix $\bb{T}_i(\omega)\in\mathbb{C}^{(N'+1)^2\times(N+1)^2}$ is, in general, not a square matrix; it transforms a field of order $N$ into a field of order $N'$. Using the property $C_{n',m'}^{n,m,q}=0$ for $|n-n'|>q$, the new \ac{SH} order $N'$ can be expressed as
\begin{equation}
	N'=N+\lceil k r_i \rceil.
	\label{eq:m8}
\end{equation}
This implies that a translation can effectively increase the \ac{SH} order of the sound field by up to $\lceil k r_i \rceil$ orders.

Note that the complexity of the calculations required to obtain $\bb{T}_i(\omega)$ is proportional to the third power of $\lceil kr_i\rceil$. This is because the upper limit of the sum in \eqref{eq:m6} depends linearly on $\lceil kr_i\rceil$ and the number of rows in $\bb{T}_i(\omega)$ increases quadratically with increasing $\lceil kr_i\rceil$. Hence, especially for large translations, the calculation of all the entries of $\bb{T}_i(\omega)$ may become relatively complicated.
Fortunately, in practice, the calculation of the translation matrix $\bb{T}_i(\omega)$ can be simplified.
Taking the first frame $i=1$ as a reference, the overall translation matrix from the reference frame to the $i^{th}$ frame, can  be expressed as
\begin{equation}
	\bb{T}_i(\omega)=\prod\limits_{l=1}^{i}\bb{T}'_l(\omega),\,\,\,i>1
	\label{eq:m8.1}
\end{equation}
where $\bb{T}'_l(\omega)$ denotes the translation matrix between the two subsequent frames $l$ and $l-1$.
Denote the translation vector between two subsequent time frames $l$ and $l-1$ by $(r'_l,\theta'_l,\phi'_l)$.
In practice, a relatively fast motion of $1$ m/s results in $kr'_l<1$ up to $5$ kHz, considering an offset of $10$ ms between the frames.
Hence, for large translations, decomposing $\bb{T}_i(\omega)$ as suggested by \eqref{eq:m8.1} has the potential to reduce the complexity of the calculations. In particular, the expression for the element of $\bb{T}'_l(\omega)$ given in \eqref{eq:m6} becomes  
\begin{align}
	\nonumber[\bb{T}'_l]_{n'^2+n'+m}^{n^2+n+m}&={j_0(k r'_l)\cdot Y_{0}^{m-m'}(\theta'_l,\phi'_l)\cdot C_{n',m'}^{n,m,0}}\\
	&+{j_1(k r'_l)\cdot Y_{1}^{m-m'}(\theta'_l,\phi'_l)\cdot C_{n',m'}^{n,m,1}}.
	\label{eq:m9}
\end{align}
Using properties of the Wigner-3j symbol it can be shown that the first term in \eqref{eq:m9} reduces to
\begin{equation}
	Y_{0}^{m-m'}(\theta'_l,\phi'_l)\cdot C_{n',m'}^{n,m,0}=\delta_{n,n'}\delta_{m,m'},
	\label{eq:m10}
\end{equation}
and for the second term it holds that  
\begin{equation}
	Y_{1}^{m-m'}(\theta'_l,\phi'_l)\cdot C_{n',m'}^{n,m,1}=0, \,\,n=n'.
	\label{eq:m11}
\end{equation}
Thus, for $kr'_l<1$, $\bb{T}'_l(\omega)$ can be calculated using the simplified form:
\begin{equation}
	\bb{T}'_l(\omega)=j_0(kr'_l)\tilde{\bb{I}}+j_1(kr'_l)\bb{C},
	\label{eq:m12}
\end{equation}
where $\tilde{\bb{I}}$ is an $(N'+1)^2\times(N+1)^2$ matrix with the entries being $1$ on the main diagonal and $0$ otherwise. Matrix $\bb{C}$ is an off-diagonal matrix with the entries given by the second term in \eqref{eq:m9}. It is interesting to note that for $r'_l\rightarrow 0$, functions $j_0(kr'_l)\rightarrow 1$ and $j_1(kr'_l)\rightarrow 0$ and $N'\rightarrow N$. This implies that for small translations the second term in \eqref{eq:m9} vanishes. Hence, consistently, $\bb{T}'_l(\omega)\rightarrow \bb{I}$. 

To summarize, a moving array model was proposed in \eqref{eq:m1}. The model accounts for motion by means of a linear transformation $\bb{W}_i(\omega)$, as detailed in sections \ref{sec:rotation} and \ref{sec:translation}.
This model is the basis for the development of the motion-aware processing methods outlined in the following section.
\section{Motion-aware processing} 
\label{sec:methods}
In the previous section, a signal model was presented that takes into account the motion of the robot. This model is utilized here to develop  methods for the processing of signals from microphone arrays installed on a moving robot. First, a method that compensates for array motion is presented; it will be referred to as the \emph{motion compensation} approach. Then, a method is presented that uses the robot motion in order to enhance the performance to a level beyond that of a stationary array. Hereafter, this will be referred to as the \emph{motion-based enhancement} approach.
\subsection{Motion compensation approach} 
\label{sec:compensated}
Suppose that $J$ consecutive samples of the \ac{STFT}, i.e., $\bb{p}(i,\omega),\,i=1,2,...,J$, were acquired using a microphone array installed on a robot. It is assumed here that during the data acquisition the sources remain at the same positions and the motion of the robot does not involve transition between different rooms.
As was mentioned above, signals from a microphone array installed on a moving humanoid robot can be processed using the \emph{stop-perceive-act} principle that demands stopping the robot during data acquisition. Following this approach and assuming that the robot is stopped during data acquisition, the \ac{PWD} function can be computed from $\bb{p}(i,\omega),\,i=1,2,...,J$ by using the stationary array model
\begin{align}
	\nonumber\hat{\bb{a}}_s(i,\omega)&=\bb{V}^{\dag}(\omega)\bb{p}_s(i,\omega)\\
	&=\bb{a}(i,\omega)+\bb{V}^{\dag}(\omega)\bb{n}(i,\omega),
	\label{eq:m13}
\end{align}
where $\bb{p}_s(i,\omega)$ denote the STFTs obtained by a stationary microphone array, as described by the model in \eqref{eq:b5}.
The major drawback of the \emph{stop-perceive-act} principle is that it imposes a behavioral constraint on the robot and, therefore, can limit the naturalness of its interaction with the surroundings.
An alternative approach, proposed in this paper, is to compensate for the motion by utilizing the moving array model presented in the previous section. Using this model, motion compensated estimation of the \ac{PWD} function can be performed from the measurements obtained by a moving array:
\begin{align}
	\nonumber\hat{\bb{a}}_m(i,\omega)&=\left[\bb{V}_i(\omega)\bb{W}_i(\omega)\right]^{\dag}\bb{p}_m(i,\omega)\\
	&=\bb{a}(i,\omega)+\left[\bb{V}_i(\omega)\bb{W}_i(\omega)\right]^{\dag}\bb{n}(i,\omega),
	\label{eq:m14}
\end{align}
where $\bb{p}_m(i,\omega)$ denote the \ac{STFT}s obtained by a moving microphone array and it is assumed that the measured noise $\bb{n}(i,\omega)$ is identical for moving and stationary arrays. It is emphasized that the transformation matrices $\bb{W}_i(\omega)$ are known; they can be calculated as explained in sections \ref{sec:rotation} and \ref{sec:translation} based on knowledge of the robot's trajectory. Note that if no compensation for motion is applied, i.e., $\bb{p}_m(i,\omega)$ is processed in accordance with \eqref{eq:m13}, then
the obtained $\hat{\bb{a}}_m(i,\omega)=\bb{W}_i(\omega)\bb{a}(i,\omega)+\bb{V}^{\dag}(\omega)\bb{n}(i,\omega),\,i=1,2,...,J$, describe different sound fields, as viewed from the array coordinate system at different points along the array trajectory. Hence, the performance of the subsequent \ac{DoA} estimation using all of $\hat{\bb{a}}_s(i,\omega),\,i=1,2,...,J$, as outlined in \emph{step 3} of the \ac{SH-MUSIC} algorithm, will be degraded. This point is further demonstrated by the simulation examples in section \ref{sec:numercompensation}.

Note that the error components that appear when using the \emph{stop-perceive-act} and the motion compensation approaches, i.e., the terms $\bb{V}^{\dag}(\omega)\bb{n}_i(\omega)$ and $\left[\bb{V}_i(\omega)\bb{W}_i(\omega)\right]^{\dag}\bb{n}_i(\omega)$, respectively, are, in general, different. Comparison of these terms can be difficult for a general $\bb{W}_i(\omega)$. 
However, for pure rotation the matrix $\bb{W}_i(\omega)=\bb{R}_i$ is unitary. Moreover, $\bb{V}_i(\omega)=\bb{V}(\omega)$, because rotation does not affect the \ac{SH} order of the field. In this case, it is straightforward to show that the power of the error terms, expressed by the trace of their covariance matrices, is identical.
For the general case, recall that the rotation and translation parts, $\bb{R}_i$ and $\bb{T}_i(\omega)$, closely approximate the identity matrix for relatively small movements between the time frames. Therefore, it may be expected that $\bb{W}_i(\omega)$ has only a small effect on the power of the error term. Thus, the motion compensation approach is expected to maintain the same performance as the \emph{stop-perceive-act} approach, while removing the constraints on the robot's behavior.

The following subsection presents a method that exploits the robot's motion in order to improve the array performance to a level beyond that of a stationary array.

\subsection{Motion-based enhancement approach} 
\label{sec:enhanced}
Section \ref{sec:moving} introduced a signal model for a moving array. Based on this model, an approach has been discussed in section \ref{sec:compensated} that compensates for the motion in the estimation of the \ac{PWD} function. It was assumed that the number of microphones and their distribution on the head surface enable estimation of the \ac{PWD} function up to the required order $N$. However, in practice, the number of microphones may be limited. See, for example, the microphone arrays of humanoid robots NAO \cite{Rump2014} and Hearbo \cite{Nakamura2013}, which have only $4$ and $8$ microphones, respectively. A small number of microphones limits the \ac{SH} order of the \ac{PWD} function that can be inferred; this limits the performance of the \ac{DoA} estimation algorithms in terms of resolution, robustness, frequency range, and the number of sources that can be localized.
In the current subsection, an approach is presented that combines the measurements from several time frames. In this way, the effective number of microphones is increased, thereby improving the performance of the subsequent \ac{DoA} estimation.

Here, it is assumed that the \ac{STFT}s of the microphone outputs were computed resulting in $I$ pressure samples at each frequency, i.e., $\bb{p}(i,\omega),\,i=1,2,...,I$. Note that the number of frames is denoted here by $I$ instead of $J$. The reason for this change of notation is discussed at the end of the section. The method presented here enables the combination of these $I$ samples to produce a single estimate of the \ac{PWD} function. 

In addition to the assumptions made in the previous section, it is assumed here that the amplitudes of the signals produced by the sources at all frequencies of interest are constant in time, i.e., the amplitudes are the same for $i=1,2,...,I$. Although this is a restrictive assumption, it appears to be relatively common in the signal processing literature \cite{Tzoreff2014,Yen1989}. In section \ref{sec:comban}, it is demonstrated that the method proposed here is reasonably robust to variations in both the amplitude and the frequency, promoting its application in practice for the localization of sources that produce quasi-periodic signals, such as fire alarms or music. 

In the case where the above assumption holds, the \ac{PWD} function $\bb{a}(i,\omega)$, as measured from the coordinate system in a reference position, differs for various values of $i$ only by the phase related to the time offset $D$ between the frames:
\begin{equation}
	\bb{a}(i,\omega)=\bb{a}(1,\omega)e^{j2\pi D\omega(i-1)/T}.
	\label{eq:c1}
\end{equation}
Note that in \eqref{eq:c1}, the reference time frame was arbitrarily chosen to be the first frame, for convenience.
By substituting \eqref{eq:c1} into the moving array model in \eqref{eq:m1}, we obtain
\begin{equation}
	\bb{p}(i,\omega)=\bb{V}_i(\omega)\bb{W}_i(\omega)\bb{a}(1,\omega)e^{j2\pi D\omega(i-1)/T}+\bb{n}(i,\omega).
	\label{eq:c2}
\end{equation}
Multiplying both sides of \eqref{eq:c2} by $e^{-j2\pi D\omega(i-1)/T}$ results in the following relation:
\begin{equation}
	\tilde{\bb{p}}(i,\omega)=\bb{V}_i(\omega)\bb{W}_i(\omega)\bb{a}(1,\omega)+\tilde{\bb{n}}(i,\omega),
	\label{eq:c3}
\end{equation}
where 
\begin{equation}
	\tilde{\bb{p}}(i,\omega)=\bb{p}(i,\omega)e^{-j2\pi D\omega(i-1)/T}
	\label{eq:c4}
\end{equation}
and
\begin{equation}
	\tilde{\bb{n}}(i,\omega)=\bb{n}(i,\omega)e^{-j2\pi D\omega(i-1)/T}.
	\label{eq:c5}
\end{equation}
The result in \eqref{eq:c3} relates the time-aligned STFTs, $\tilde{\bb{p}}(i,\omega)$, for $i=1,2,...,I$, to the same \ac{PWD} function $\bb{a}(1,\omega)$ that was captured by the array during the reference (first) time frame. Next, by concatenating the time-aligned STFTs into a single measurement vector, a combined system can be constructed:
\begin{equation}
	\tilde{\bb{p}}(\omega)=\bb{A}(\omega)\bb{a}(1,\omega)+\tilde{\bb{n}}(\omega),
	\label{eq:c6}
\end{equation}
where $\tilde{\bb{p}}(\omega)=[\tilde{\bb{p}}^T(1,\omega),\tilde{\bb{p}}^T(2,\omega),...,\tilde{\bb{p}}^T(I,\omega)]^T$ and $\tilde{\bb{n}}(\omega)=[\tilde{\bb{n}}^T(1,\omega),\tilde{\bb{n}}^T(2,\omega),...,\tilde{\bb{n}}^T(I,\omega)]^T$ are column vectors with the dimensions $I\cdot M\times 1$.
Matrix $\bb{A}(\omega)$ is given by
\begin{equation}
	\bb{A}(\omega)=
	\left(
	\begin{tabular}{c}
		$\bb{V}_1(\omega)\bb{W}_1(\omega)$\\
		$\bb{V}_2(\omega)\bb{W}_2(\omega)$\\
		$\vdots$\\
		$\bb{V}_I(\omega)\bb{W}_I(\omega)$
	\end{tabular}
	\right) \in \mathbb{C}^{I\cdot M\times (N+1)^2},
\label{eq:c7}
\end{equation}
where $\bb{W}_1=\bb{I}$ is the identity matrix; this is because the first frame was chosen to serve as the reference. Using the model in \eqref{eq:c6}, the \ac{PWD} function in the reference frame can be obtained by applying the pseudo inverse of the combined matrix, i.e.,
\begin{equation}
	\hat{\bb{a}}(1,\omega)=\bb{A}^{\dagger}(\omega)\tilde{\bb{p}}(\omega).
	\label{eq:estimate_pwd_combined}
\end{equation}
Note that the motion-based enhancement approach described here provides a single estimate of the \ac{PWD} function using $I$ samples of the sound pressure. Thus, in order to obtain reliable estimates of the covariance matrices, an extended data acquisition time of $J\cdot I$ frames may be required, as compared to $J$ frames that are required in the motion compensation approach. 
The advantage of the motion-based enhancement method is that it enables us to estimate $(N+1)^2$ \ac{SH} coefficients of the \ac{PWD} function by jointly using $I\cdot M$ equations, as compared to only $M$ equations available when using the motion compensation method. Hence, when using the motion-based enhancement approach, the \ac{SH} order $N$ of the estimated \ac{PWD} function can be increased. This is discussed in more detail in the next subsection.

\subsection{Analysis of the motion-based enhancement approach} 
\label{sec:comban}
The motion-based enhancement approach presented in \eqref{eq:estimate_pwd_combined} enables the estimation of the \ac{PWD} function by jointly using the information gathered by the array from $I$ different time frames, which, due to the array motion, were obtained at different array positions.
The apparent advantage of this approach over the motion compensation approach presented in \eqref{eq:m14} is that the number of equations it provides for the estimation of the \ac{PWD} function is $I$ times greater. This is particularly true for relatively large movements for which $\bb{W}_i(\omega)$ significantly differ from each other. On the other hand, for small movements, matrices $\bb{W}_i(\omega)\rightarrow \bb{I}$, and $\bb{A}(\omega)\rightarrow [\bb{V}(\omega)^T\,\bb{V}(\omega)^T\,\cdots\,\bb{V}(\omega)^T]^T$, implying that no independent equations are added.
Hence, it is important to gain an insight into the degree to which the motion-based enhancement approach can improve the estimation of the \ac{PWD} function, and into the way in which the improvement depends on the array speed, the trajectory, and the frequency range.

The effective increase in the number of independent equations in \eqref{eq:c6} is assessed here through the \emph{effective rank} of $\bb{A}(\omega)$ \cite{Roy2007}. This quantity is based on the uniformity of the singular values of a matrix and, as opposed to the discrete values of the actual matrix rank, it provides a continuous estimate of the effective dimension of a system. This property makes it suitable for assessing the dimension of a system as a function of continuous parameters such as frequency, radius of translation, and angle of rotation \cite{Tourbabin2012}. In addition, this measure was shown to be related to array beamforming and \ac{DoA} estimation performance \cite{Tourbabin2014Dec}, making it particularly suitable for the analysis presented here.
First, the effective rank of $\bb{A}(\omega)$, combining $I=2$ frames, was calculated as a function of the array displacement between the time frames.  
For simplicity, we used a rigid equiangular spherical array \cite{Rafaely2005} of $13$ microphones distributed with a $45^{\circ}$ spacing in elevation and a $90^{\circ}$ spacing in azimuth.
The array radius is $r_a=6$ cm and the \ac{SH} order of the field is assumed to be limited to $N=6$, implying that the number of columns in $\bb{A}(\omega)$ is $(N+1)^2=49$.
In order to compute $\bb{A}(\omega)$, the steering matrix $\bb{V}(\omega)$ is required (see \eqref{eq:b6}). For a rigid spherical array, this matrix can be computed using an analytic expression \cite{Rafaely2005}. The rotation and translation matrices were calculated using \eqref{eq:m3} and \eqref{eq:m6}.

\begin{figure}[ht] 
	\centering
	\begin{tabular}{cc}
		\hspace{-7pt}
		\includegraphics[width=0.48\columnwidth]{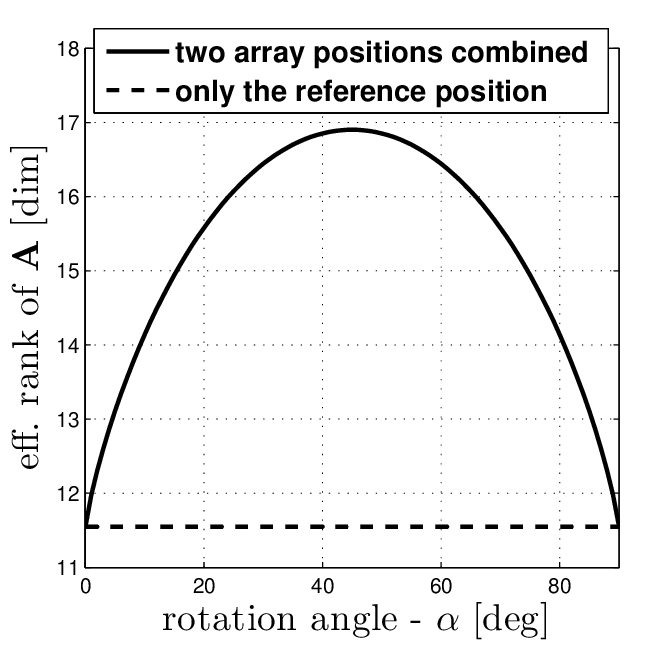} &
		\hspace{-13pt}
		\includegraphics[width=0.48\columnwidth]{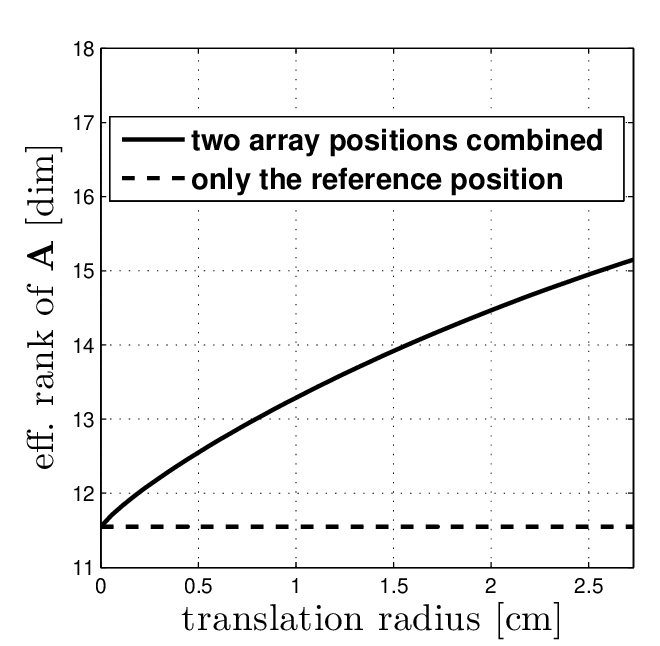}\\
		(a) & (b)
	\end{tabular}
	\caption{Effective rank of the combined model matrix $\bb{A}(\omega)$ at $2$ kHz (see \eqref{eq:c6}) as a function of (a) rotation
	angle $\alpha$, (b) translation distance $r$.}
	\label{fig:c1}
\end{figure}

Two different modes of motion were considered: (a) rotation about the $z$ axis by an angle $\alpha$ and (b) translation in the direction $(90^{\circ},90^{\circ})$ by a distance $r$. The effective rank of $\bb{A}(\omega)$ at $2$ kHz as a function of $\alpha$ and of $r$ is presented in Fig. \ref{fig:c1}. The effective rank of the array in only the reference position is also plotted in order to provide a reference value. 
It can be seen that the effective rank of $\bb{A}(\omega)$ increases significantly for larger displacements (either rotation or translation).
The increase in the effective rank is consistent with the fact that increasing the displacement is similar to increasing the effective array aperture. Therefore, it is expected to increase the spatial information about the surrounding sound field gathered by the array. 
In the rotation case, the improvement is maximal at $\alpha=45^{\circ}$. This is due to the spatial symmetry of the rotations by $\alpha$ and $90^{\circ}-\alpha$ for this particular array geometry.
Note that combining two time frames may be thought of as doubling the array to obtain $26$ virtual microphones. This has the potential to increase the rank of $\bb{A}$ to $26$. However, in the configuration considered here, the rank can be effectively increased to about $17$ dimensions, as it is demonstrated by Fig. \ref{fig:c1}.a.

\begin{figure}[ht] 
	\centering
	\includegraphics[width=\columnwidth]{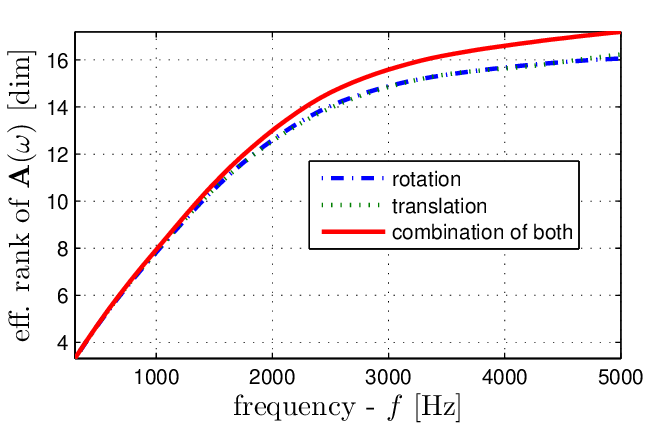}
	\vskip-13pt
	\caption{Effective rank of $\bb{A}(\omega)$ as a function of frequency for three different
	 modes of motion. The rotation and translation curves almost overlap.}
	\label{fig:c2}
\end{figure}

The increase in the effective rank for a particular displacement may depend on frequency.
Fig. \ref{fig:c2} provides an example of the dependence of the effective rank of $\bb{A}(\omega)$ on frequency. Three different modes of motion are considered: (i) rotation by $\alpha=3^{\circ}$, (ii) translation by $5$ mm, and (iii) a combination of both, with the rotation followed by the translation. 
It can be seen that in all three cases the effective rank increases with frequency.
Recall that the rotation matrix $\bb{R}_i$ does not depend on frequency. Thus, the dependence on frequency of $\bb{A}(\omega)$ that results in the increase of its effective rank is due to the frequency dependence of $\bb{V}(\omega)$.
Note that, for this particular set of parameters, the improvements in the effective rank due to the rotation or the translation alone are similar. The effect of the complex movement (rotation followed by translation) is greater than that of the rotation or the translation alone.

To summarize, the effective rank of the combined system matrix $\bb{A}(\omega)$ increases for larger displacements and higher frequencies.
Thus, the benefit of combining the different frames is expected to be more pronounced with an increase in the speed of the robot motion and the processing frequency range. 
This tendency is expected to be similar for different types of motion, including rotation, translation, and a combination of both. 
Hence, for conciseness, the investigation in the following sections considers a typical example of a common robot head motion \-- rotation about the $z$ axis. This type of motion in humanoid robots represents, for example, left and right head rotations during a conversation with multiple speakers, steering of the head in response to a sudden acoustic event, and scanning of the surroundings.

\section{Simulation study: motion compensation} 
\label{sec:numercompensation}
This section presents an investigation of the motion compensation method. This investigation demonstrates the \ac{DoA} estimation error induced by array motion with realistic parameters and the ability of the motion compensation method to reduce this error. The investigation is based on simulated microphone outputs from a moving microphone array and on the \ac{SH-MUSIC} algorithm. An investigation of the motion-based enhancement approach is presented in the next section. Note that in both sections \ref{sec:numercompensation} and \ref{sec:numerenhancement}, the effect of the robot's torso on the sound propagation is neglected at present. However, the effect may be significant and may depend on the posture of the robot. A study of the appropriate adjustments of the \ac{HRTF} \cite{Algazi2002} that would be required to take this effect into account is suggested for future work. Nevertheless, section VII demonstrates experimentally that \ac{DoA} estimation at high frequency may be robust to this effect. 

\subsection{Simulated setup} 
\begin{figure}[ht] 
	\centering
	\includegraphics[width=\columnwidth]{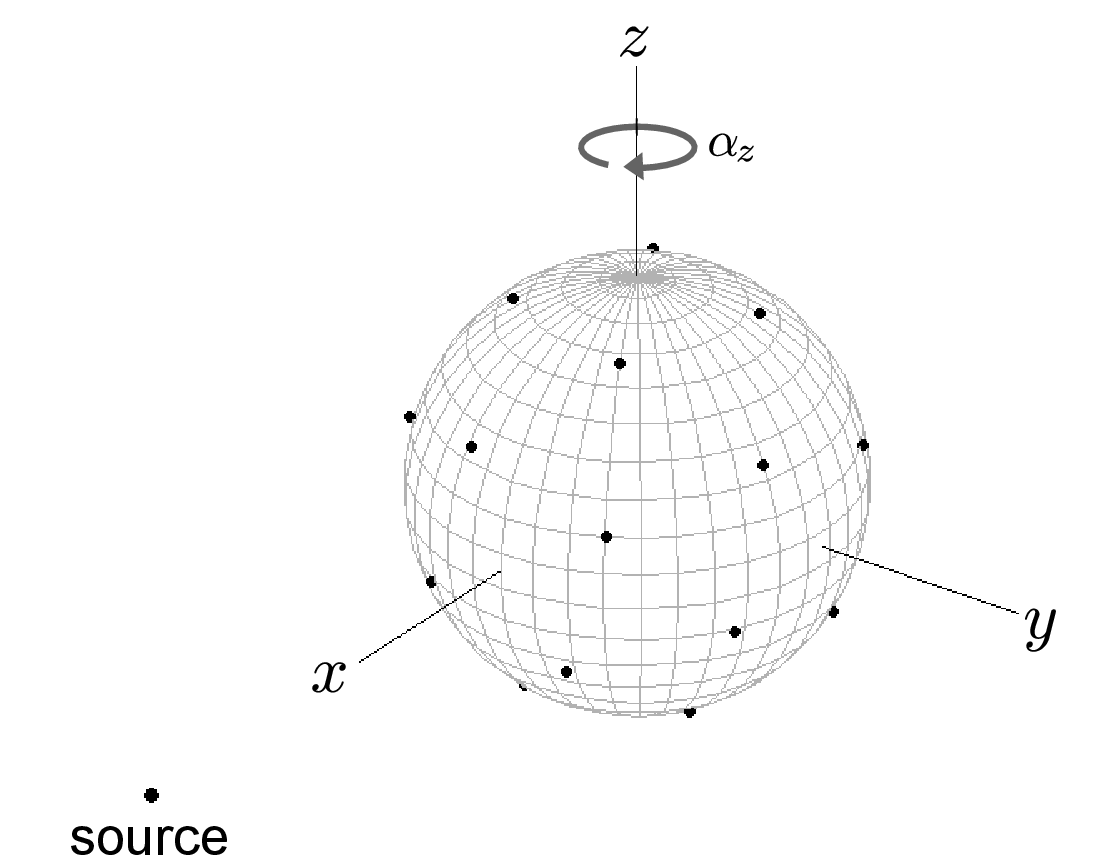}
	\caption{A schematic illustration of the microphone array, the initial source position, and
	 the rotation direction that were used for the investigation of \ac{DoA} estimation
	 performance using the motion compensation approach. Microphone positions are indicated
	 by the black dots on the sphere surface.}
	\label{fig:e1}
\end{figure}

The investigation in this section uses a simulated rigid spherical array \cite{Rafaely2005}. 
The array radius is $6$ cm. The array consists of $24$ nearly uniformly \cite{Hardin1996} distributed microphones. This array enables the estimation of the \ac{PWD} function up to $3$ kHz \cite{Rafaely2005}. As noted above, the investigation is based on a relatively common head motion mode \-- rotation about the $z$ axis with constant angular velocity denoted by $\alpha_z$. A single source in a free field was simulated; it was initially positioned in the direction $(\theta,\phi)=(\pi/2,0)$ in the array coordinate system. The array, the source, and the rotation direction are illustrated schematically in Fig. \ref{fig:e1}.
An example involving the head geometry of a real humanoid is presented in section \ref{sec:numerenhancement}.
In addition, the investigation in this section assumes that the source and the array are positioned in a free field in order to simplify the simulation and the interpretation of the results. A more realistic acoustic scenario may include room reverberation, which is expected to have a significant effect on \ac{DoA} estimation performance. However, it should be emphasized that the motion compensation method presented here only aims to remove the effect of motion from the \ac{PWD} estimates. Hence, the degree to which reverberation affects performance is expected to be determined solely by the robustness of the \ac{DoA} estimation algorithm that uses these \ac{PWD} estimates \cite{Nadiri2014}.

The outputs of all $24$ microphones were simulated by filtering the source signal with the frequency responses of individual microphones. The responses were obtained analytically by using the expression for the array steering vector in the \ac{SH} domain \cite{Rafaely2005}.
The filtering was carried out using the overlap-save technique. The rotation was simulated by changing the angle between the source and the array and updating the filter each millisecond. Considering the maximum angular velocity of $180$ deg/s that was used in the simulations, updating the filter each millisecond corresponds to a spatial resolution of $0.18$ deg. A speech signal from the TIMIT database \cite{Garofolo1993} downsampled to $10$ kHz was used as a source signal.
Prior to the \ac{DoA} estimation, white Gaussian noise with appropriate wideband \ac{SNR} levels was added to the microphone outputs in order to simulate the additive noise.

\subsection{\ac{DoA} estimation parameters and performance measures} 
The \ac{DoA} estimation was performed using the \ac{SH-MUSIC} algorithm described in section \ref{sec:music}. The \ac{STFT} in \emph{step 1} of the algorithm was calculated using Hamming windows of length $256$ samples with $50\%$ overlap.
A block of $60$ consecutive frames was used for the estimation of the covariance matrix in $\emph{step 3}$. Hence, the overall data acquisition time required to produce a \ac{DoA} estimate was about $0.75$ s.
Note that the array used here enables aliasing-free estimation of the \ac{PWD} function up to the \ac{SH} order of $N=3$. The \ac{PWD} function of this order was estimated in the frequency range of $1800-2700$ Hz. The lower frequency limit ensures that the simulated sound pressure contains significant energy from the required \ac{SH} order. 

The performance of the algorithm was assessed by observing the \ac{STD} and the average of the \ac{DoA} estimation error angle $\Delta$, which is defined as the angle between the true and the estimated \ac{DoA}. The \ac{STD} and the average of the error were calculated using $60$ consecutive \ac{DoA} estimation trials that differed by the noise component.

\subsection{Results and discussion} 
\begin{figure}[ht] 
	\centering
	\includegraphics[width=\columnwidth]{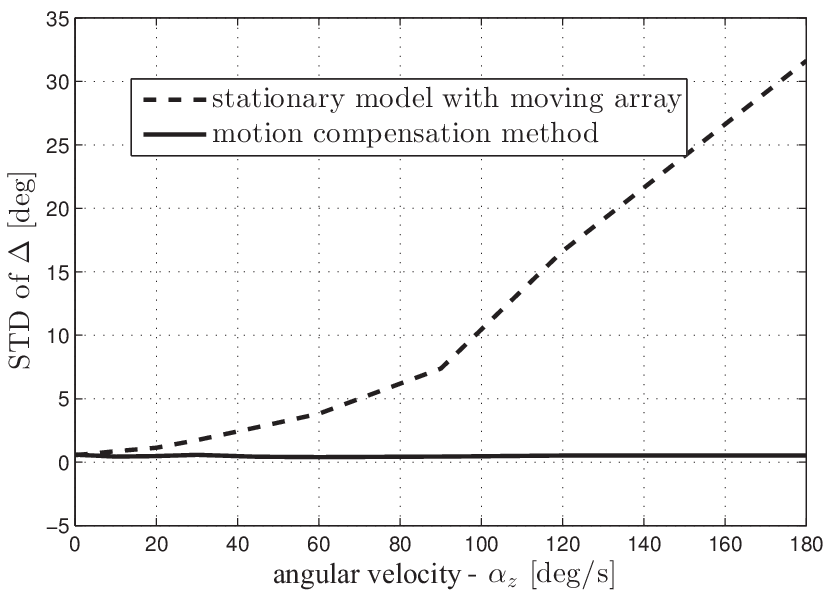}
	\caption{STD of the \ac{DoA} estimation error $\Delta$ as a function of angular velocity $\alpha_z$.}
	\label{fig:e2}
\end{figure}

\begin{figure}[ht] 
	\centering
	\includegraphics[width=\columnwidth]{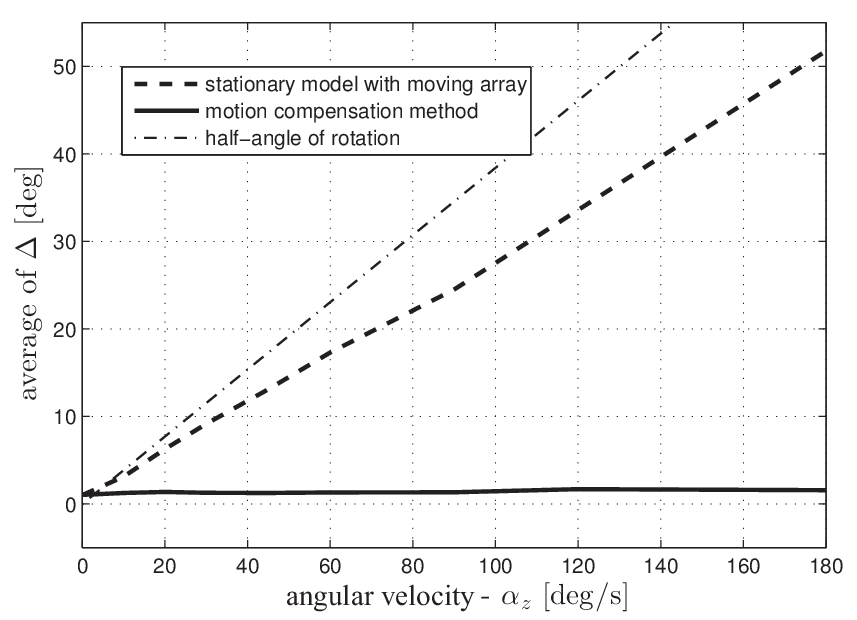}
	\caption{Average of the \ac{DoA} estimation error $\Delta$ as a function of angular velocity $\alpha_z$. The curve denoted
	``half-angle of rotation'' displays half of the angle that the array covers during the acquisition of the block of $60$ STFT
	frames required to produce a single \ac{DoA} estimate. See text for the details.}
	\label{fig:e3}
\end{figure}

Here, we examine the degree to which the \ac{DoA} estimation performance is degraded when motion compensation is not applied to a moving array. For this purpose two different ways of estimating the \ac{PWD} function are compared: (i) using the motion compensation method and (ii) using the stationary array model, in spite of the fact that the array is moving.
The \ac{STD} and the average error values obtained using the two methods are presented in Figs. \ref{fig:e2} and \ref{fig:e3} as a function of the array angular velocity $\alpha_z$.  These results were obtained at an \ac{SNR} of $10$ dB.
 
It can be seen that when no compensation for the motion is employed, both the \ac{DoA} estimation variance and the bias increase with increasing angular velocity. This is due to the fact that during the data acquisition the source position relative to the array changes continuously, introducing an error into the estimated \ac{PWD} function. 
It is interesting to note that, especially at lower angular velocities, the bias in the \ac{DoA} estimation roughly follows the half-angle of rotation covered during the data acquisition, calculated as $\alpha_z\cdot0.75/2$ [deg] (see Fig. \ref{fig:e3}). This is due to the averaging of the \ac{PWD} function samples obtained at different array positions when estimating the covariance matrix in \eqref{eq:b8}. Hence, at least for rotation, the bias can be partially corrected by a simple addition of half of the covered angle.
Note that for $\alpha_z=0$ deg, Figs. \ref{fig:e2} and \ref{fig:e3} display the performance of a stationary array.
As expected, when motion compensation is employed this level of performance is maintained for arrays rotating at all $\alpha_z$ values in the range.

\section{Simulation study: motion-based enhancement} 
\label{sec:numerenhancement}
The previous section presented an investigation of the motion compensation approach using an array with a relatively large number of microphones. In practice, the arrays installed in humanoid robots may have a limited number of microphones. This motivated the development of the motion-based enhancement approach described in section \ref{sec:enhanced}, which proposes to combine the observations from different time frames (see \eqref{eq:c6}) for the estimation of the \ac{PWD} function in \emph{step 2} of the \ac{SH-MUSIC} algorithm. The ability of this approach to increase the information gathered by the array was analyzed in section \ref{sec:comban}. The current section demonstrates that the approach is capable of improving actual \ac{DoA} estimation performance under realistic conditions.

\subsection{Simulated setup} 
\begin{figure}[ht] 
	\centering
	\includegraphics[width=\columnwidth]{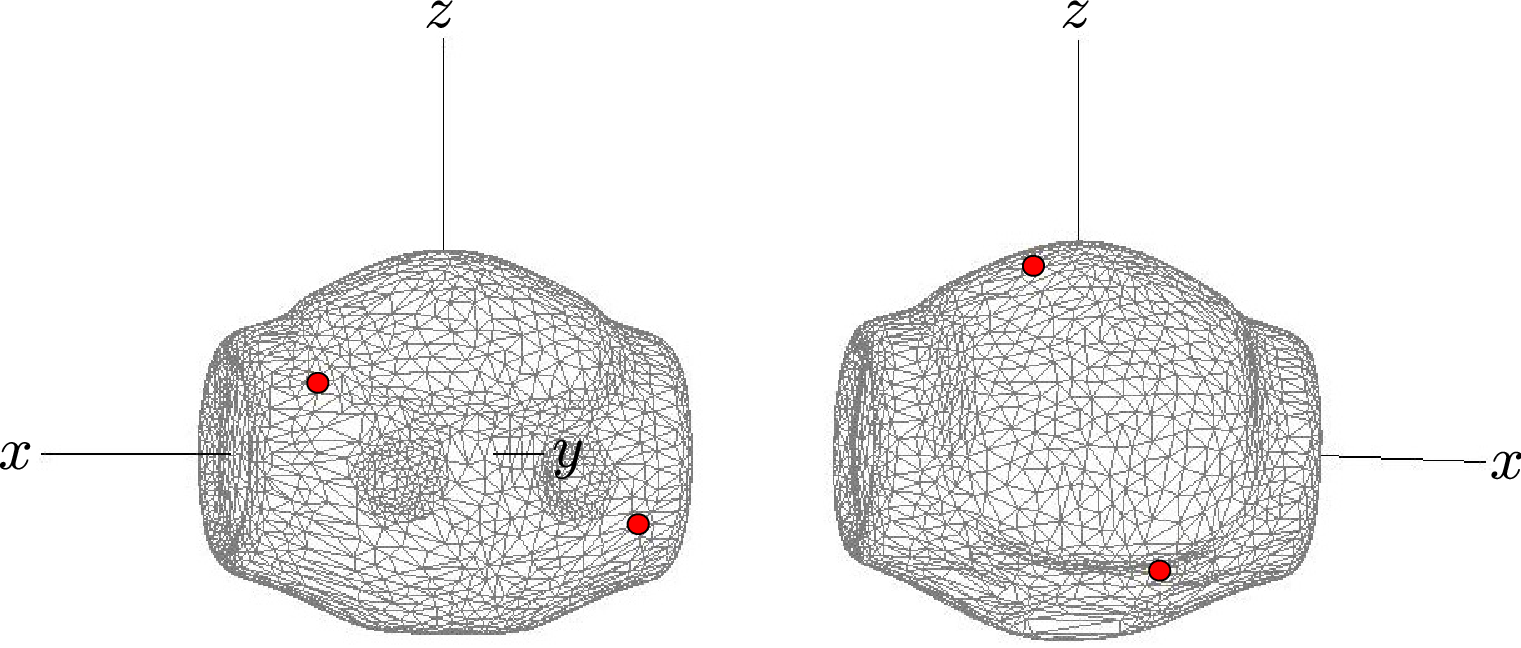}
	\caption{Schematic illustration of the $4$-microphone array used for the demonstration of
	the performance of the motion-based enhancement approach.}
	\label{fig:e5}
\end{figure}

The study in this section is based on the head of the existing humanoid robot NAO \cite{Rump2014}, with an average radius of about $6.25$ cm. This radius is similar to the array used in the previous section. However, here, the number of microphones is limited to only $M=4$, which is the case in the existing Phase-I and Phase-II models of this robot.
The geometry of the head surface and the distribution of microphones used here are schematically illustrated in Fig. \ref{fig:e5}. 
As in the previous section, this study also focuses on array rotation about the $z$ axis with a constant angular velocity $\alpha_z$. 
Recall that the motion-based enhancement approach assumes a periodic signal, i.e., a signal with a constant amplitude at a given frequency, at least over the sound acquisition time.
It was pointed out that, in practice, a periodic signal may be produced by an alarm or a music source. Hence, in this section, we use a pure tone source of frequency $3100$ Hz, imitating a fire alarm system \cite{NFPA2007}.

Microphone outputs were simulated using the overlap-save filtering technique, as before. The array steering vectors used in the filtering procedure were obtained by means of a  \ac{BEM} simulation that is based on the geometry of the head \cite{Tourbabin2014Dec}. 
One source direction was simulated at a time. In total, $20$ nearly-uniformly distributed source directions \cite{Hardin1996} were simulated. The \ac{DoA} estimation performance described below represents the average over all simulated directions. 
An additive noise component was simulated by adding white Gaussian noise with various narrowband \ac{SNR} levels.

\subsection{\ac{DoA} estimation parameters} 
\ac{DoA} estimation was performed using the \ac{SH-MUSIC} algorithm. The algorithm was based on the STFT with the same parameters as in section \ref{sec:numercompensation}.
A total of $180$ time frames were used in order to produce a single \ac{DoA} estimate, resulting in an acquisition time of $2.3$ seconds.
Estimation of the \ac{PWD} function was based on the combined model in \eqref{eq:c6} with various numbers of the combined frames $I$.
The narrowband covariance matrix $\bb{Q}(\omega)$ in \emph{step 3} of the algorithm was estimated using $J=\lfloor 180/I \rfloor$ consecutive estimates of the \ac{PWD} function. For example, for $I=30$, the covariance matrix was estimated by averaging over $6$ \ac{PWD} estimates, i.e., $i=1,2,...,6$. 

Recall that, as demonstrated in section \ref{sec:comban}, the effective rank of the combined steering matrix $\bb{A}(\omega)$ can be significantly lower than the actual rank of the matrix. Therefore, some of its singular values can be close to zero. Hence, for robustness purposes, the pseudo-inverse in \eqref{eq:estimate_pwd_combined} was calculated using the Singular Value Decomposition approach \cite{Golub1996}, while inverting only the singular values greater than $1/3$ of the largest singular value.

\subsection{Results and discussion}
\label{sec:me_frames_velocity}
Recall that the array simulated in this section is configured around a humanoid head with an average array radius of $6.25$ cm. This array is used to process a sound field at the frequency of $3100$ Hz. These parameters define the effective \ac{SH} order of the sound field, which is $\lceil kr\rceil=4$.
At the same time, the array used here consists of only $4$ microphones implying that without combining frames, as dictated by the motion-based enhancement approach, the array is capable of aliasing-free estimation of the \ac{PWD} function of only the first order. Therefore, severe aliasing is expected to reduce the \ac{DoA} estimation performance.
Nevertheless, using the motion-based enhancement approach and increasing the number of combined frames is expected to improve the effective rank of $\bb{A}(\omega)$. This allows us to increase the estimated \ac{SH} order of the \ac{PWD} function in \eqref{eq:estimate_pwd_combined} to $N=4$, thereby achieving the effective \ac{SH} order of the field.

\begin{figure}[ht] 
	\centering
	\includegraphics[width=\columnwidth]{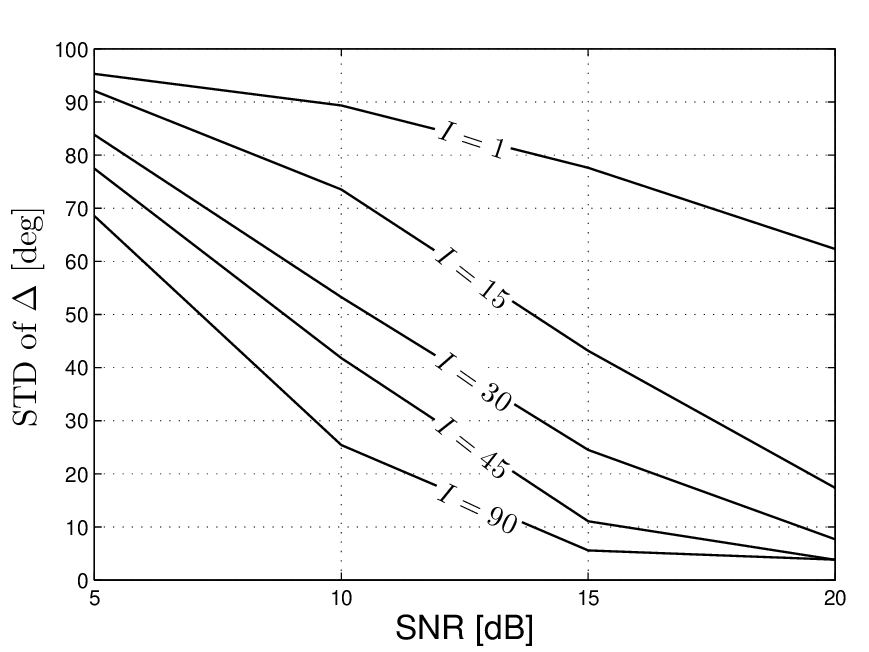}
	\caption{Performance of the \ac{SH-MUSIC} \ac{DoA} estimation algorithm using the motion-based
	 enhancement approach as a function of the number of combined frames $I$, with constant angular velocity
	  $\alpha_z=180$ deg/s.}
	\label{fig:e6}
\end{figure}

\begin{figure}[ht] 
	\centering
	\includegraphics[width=\columnwidth]{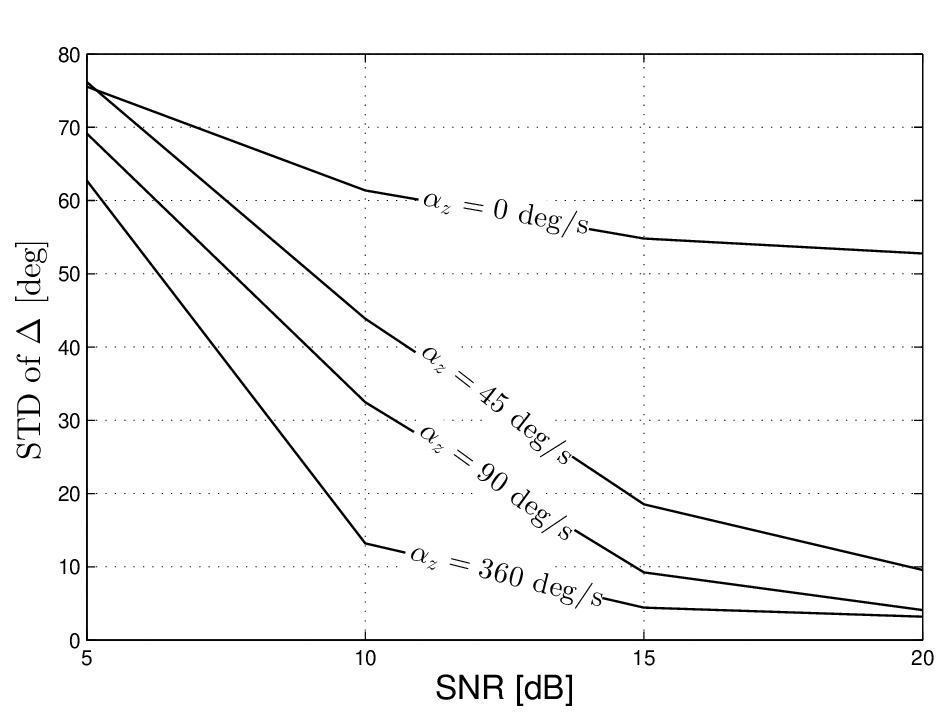}
	\caption{Performance of the \ac{SH-MUSIC} \ac{DoA} estimation algorithm using the motion-based
	enhancement approach for different values of the array angular velocity $\alpha_z$ and a fixed
	number of combined frames, $I=90$.}
	\label{fig:e7}
\end{figure}

The \ac{DoA} estimation performance as a function of the \ac{SNR} and for different numbers of combined frames $I$ is presented in Fig. \ref{fig:e6}. 
The reduced \ac{DoA} estimation accuracy for $I=1$ is, as expected, due to the detrimental effect of spatial aliasing. Nevertheless, increasing the number of combined frames, consistently increases the effective rank and the number of significant singular values of $\bb{A}(\omega)$, as summarized in Table \ref{tab:sv_I}. 
Note that the maximum effective rank obtained in this simulation is $19.6$ dimensions, which does not realize the full potential of the $25$ columns of $\bb{A}(\omega)$. 
Nonetheless, the increase in the effective rank is significant enough to produce a substantial improvement in the \ac{DoA} estimation accuracy, as demonstrated in Fig. \ref{fig:e6} for $I=15,30,45$, and $90$. 
A similar improvement is obtained when increasing the angular velocity, $\alpha_z$, while keeping the same number of combined frames, as demonstrated in Fig. \ref{fig:e7}.
This is because increasing $\alpha_z$ for a given $I$ increases the effective array aperture, which, in turn, increases the effective rank of $\bb{A}(\omega)$. Increasing the effective rank leads to improved \ac{DoA} estimation performance, as before.
\begin{table}[ht]
	\centering
	\caption{The effective rank and the number of significant singular values of $\bb{A}
	(\omega)$ for various numbers of combined frames $I$ at frequency $3100$ Hz and angular
	velocity $\alpha_z=180$ deg/s.}
	\begin{tabular}{llllll}
		\hline
		$I$ & $1$ & $15$ & $30$ & $45$ & $90$  \\
		\hline
		eff. rank of $\bb{A}(\omega)$ & $4.0$ & $7.8$ & $10.8$  & $13.4$  & $19.6$ \\
		norm. sing. val. $>1/3$ & $4$ & $5$ & $8$ & $10$ & $17$\\
		\hline
	\end{tabular}
	\label{tab:sv_I}
\end{table}

Recall that the motion-based enhancement method assumes that the amplitude of the source does not change from frame to frame during a sound acquisition time period.  Hence, the above analysis was carried out using a pure tone source. In order to assess the sensitivity of the method to deviations from this constraint, two variations of the pure tone were considered: \ac{AM} and \ac{FM}. Note that both variations will affect the instantaneous amplitude of the source component in the \ac{STFT} domain, which explicitly violates the above assumption.

\begin{figure}[ht]
	\centering
	\includegraphics[width=\columnwidth]{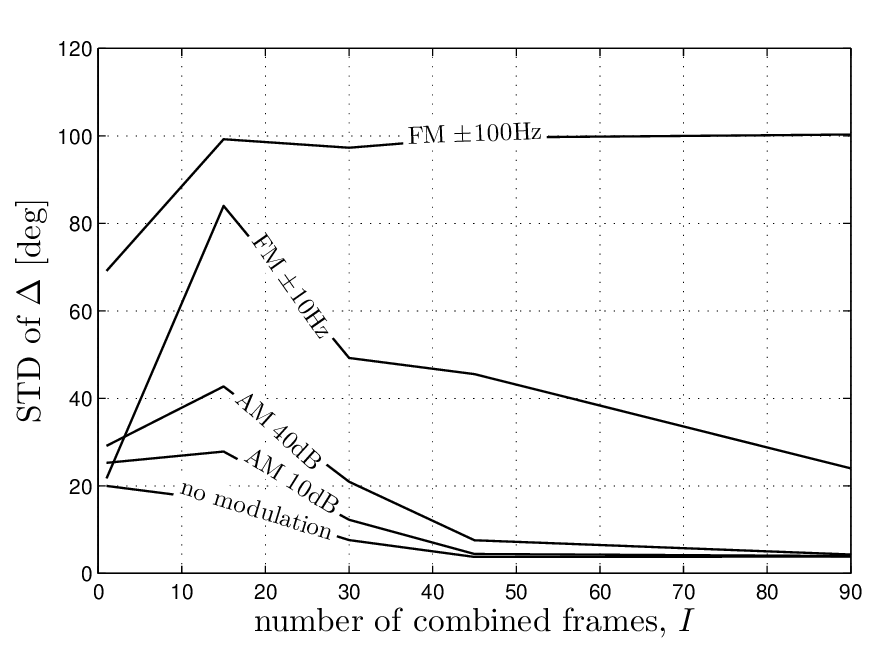}	
	\caption{Analysis of the effect of the source amplitude variation on the performance of the motion-based enhancement method. For the
	 \ac{AM} case, the ratio between the maximum and minimum amplitudes is specified in dB. For the \ac{FM} case, the deviation from the
	 tone frequency is specified in Hz. In both cases, the modulation sine wave had a frequency of $3$ Hz, completing a cycle in about
	 $13$ time frames. The results were obtained under a \ac{SNR} level of $20$ dB.}
	\label{fig:e8}
\end{figure}

The performance of the motion-based enhancement method for various levels of \ac{AM} and \ac{FM} modulations is plotted in Fig. \ref{fig:e8}.
It can be seen that increasing the amount of variation either in amplitude or in frequency has a significant effect on the performance, mainly for low values of $I$. The method seems to be more robust to variations in the amplitude of the source than to the variations in frequency, especially for large $I$. The sensitivity to variations in frequency can be attributed to the fact that a sufficiently large change in the instantaneous frequency causes most of the signal energy to shift to the neighbouring frequency bins, thereby significantly reducing the \ac{SNR} in the frequency bins that are used for the \ac{DoA} estimation. A more detailed investigation of the robustness of the method to amplitude and frequency variations is proposed for future work.

Recall that the simulations in this section were carried out assuming sources in a free field. In practice, reverberation may reduce the performance of the motion-based enhancement method. In order to improve robustness to reverberation, the method can be combined with the \ac{DPD} test \cite{Nadiri2014}. The investigation of this approach is out of the scope of the current paper and is left for future work.

\section{Experimental Study} 
\label{sec:experiment}
In the current section, the processing methods proposed in this paper are applied to experimentally obtained data, based on the full-body humanoid robot NAO.  
Recall that a mismatch between the head-only based numerically-simulated steering vectors and the true array steering vectors is therefore expected in this case. Imprecise geometric modelling of the head and imprecise microphone positioning may also contribute to the expected mismatch in the steering vectors. The main purpose of the current section is, therefore, to demonstrate that the proposed methods are robust to these and other real-world related mismatches.

\begin{figure}[ht] 
	\centering
	\includegraphics[width=\columnwidth]{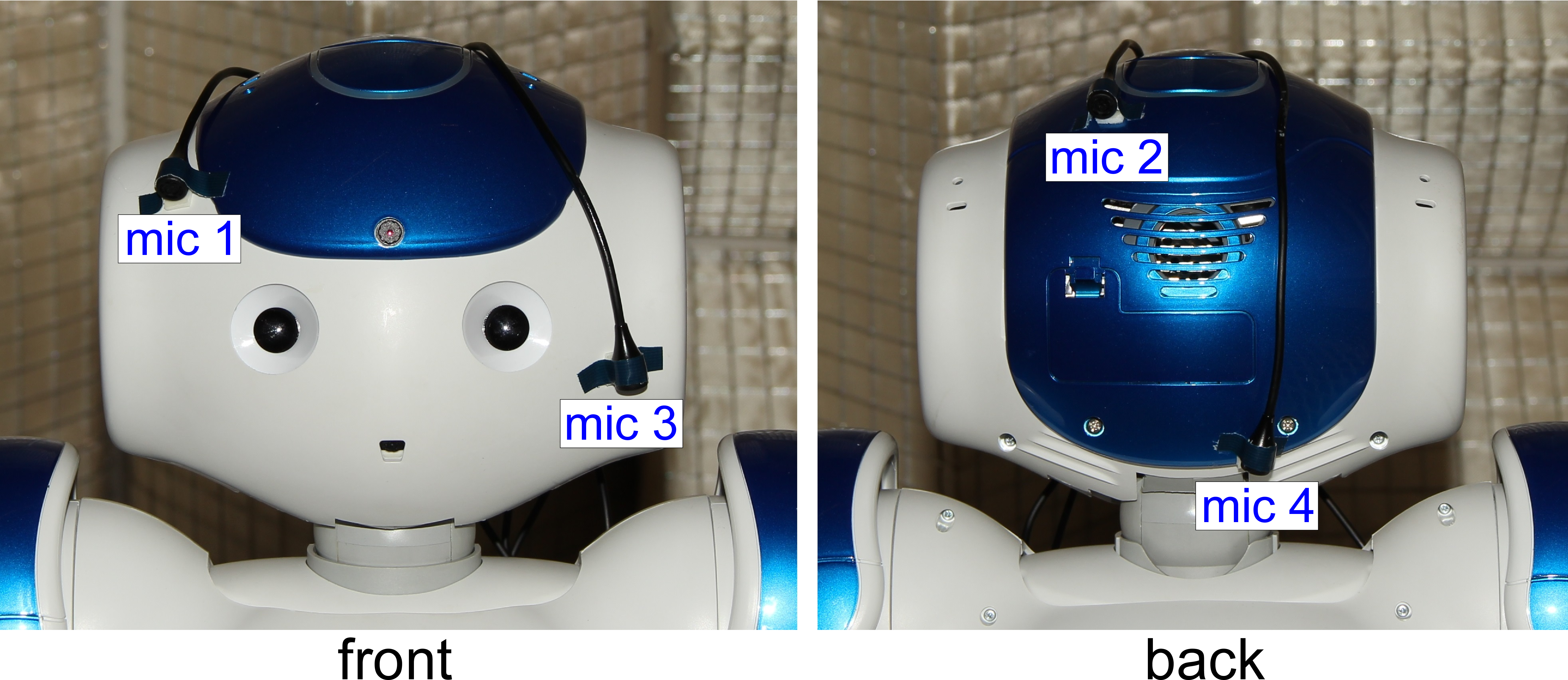}
	\caption{Positioning of the microphones on the head of the humanoid robot NAO used in the experiment.}
	\label{fig:realarray}
\end{figure}

The experiment was performed in an anechoic chamber with dimensions of $2\times 2\times 2$ m. A loudspeaker (KRK systems, Rokit 6) was set to produce a tone at a frequency of $3100$ Hz. The humanoid robot NAO was positioned in the chamber facing the loudspeaker. The head of the robot was set to rotate from left to right with constant angular velocity of $180$ deg/s. An array of four microphones (AKG, C417PP) was positioned on the head of the robot as illustrated in Fig. \ref{fig:realarray}. The positions of the microphones were chosen to be as close as possible to those used in the simulations in the previous section. The signals picked by the microphones were fed into a multichannel sound card (Focusrite, Scarlett 18i20) and sampled at a $10$ kHz sampling rate. The methods used in sections \ref{sec:numercompensation} and \ref{sec:numerenhancement} with the same parameters were applied to the obtained microphone signals. 

\begin{figure}[ht] 
	\centering
	\includegraphics[width=\imgwidth\columnwidth]{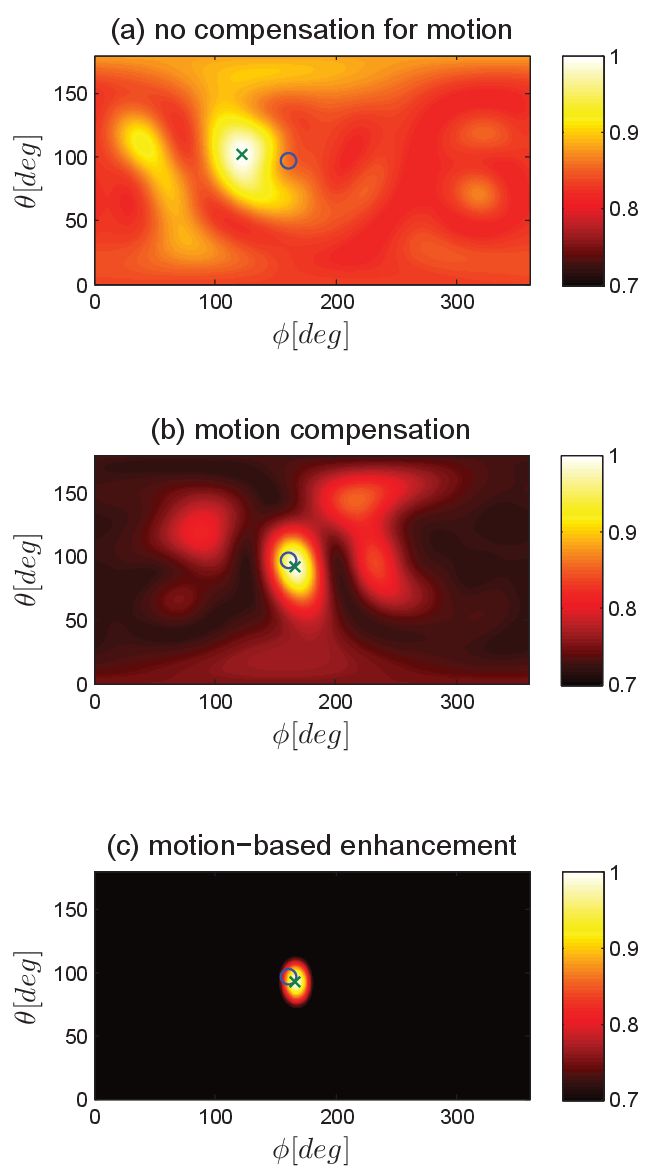}
	\caption{Spatial spectra of the \ac{SH-MUSIC} algorithm applied to the \ac{PWD} function estimated from the experimental data using
	 three different methods: (a) no compensation for motion, (b) with motion compensation, and (c) motion-based enhancement with $I=35$
	 frames. The ground-truth \ac{DoA} is indicated by $\circ$, the estimated \ac{DoA} is indicated by $\times$.}
	\label{fig:spectra}
\end{figure}

Figure \ref{fig:spectra} shows spectra of the \ac{SH-MUSIC} algorithm applied to the estimated \ac{PWD} function using three different methods: (a) no compensation for motion, (b) with motion compensation, and (c) motion-based enhancement with $I=35$ frames.
It can be seen that when no compensation for motion is applied (Fig. \ref{fig:spectra}.a), the estimated \ac{DoA} lags behind the true \ac{DoA} in the direction of rotation by about $40$ deg. Moreover, the spectrum in this case contains numerous additional peaks which may further reduce the performance of the estimation. When the motion compensation method is applied (Fig. \ref{fig:spectra}.b), the estimated \ac{DoA} is much closer to the true \ac{DoA}. However, the spectrum still contains several additional peaks, due to spatial aliasing. The motion-based enhancement method (Fig. \ref{fig:spectra}.c) overcomes the effect of aliasing and produces a much cleaner spatial spectrum, as compared to both previous cases. The spectrum contains a single peak in the vicinity of the true \ac{DoA}, as expected in the case of a single source in an anechoic chamber.
\section{Conclusion} 
\label{sec:conclusion}

In this paper, an approach was developed for \ac{DoA} estimation using microphone arrays installed on moving humanoid robots. A signal model was presented that can account for the motion of the robot using the \ac{SH} domain. Based on this model, two different processing methods were proposed. The first is the \emph{motion compensation} method, which was shown to reduce the motion-related error in the \ac{DoA} estimation when applied to the data acquired by a moving robot. The second is the \emph{motion-based enhancement} method. This method was shown to exploit the motion of the robot to improve the \ac{DoA} estimation performance to a level beyond that of a stationary array. Future work may include the extension of the \emph{motion-based enhancement} method to address non-periodic sources and reverberation. 

\section*{Acknowledgment} 
The research leading to these results has received funding from the European Union's Seventh Framework Programme (FP7/2007-2013) under grant agreement no. 609465.

\appendix[Proof of the unitarity of the rotation matrix] 
Although rotation matrices such as those that belong to $SO(3)$ are orthonormal, rotation in this paper is performed directly in the SH domain.  Therefore, for completeness, it is shown in this appendix that the Wigned-D matrices used for such rotation are, indeed, unitary.

A general rotation Wigner-D matrix of order $N$ is denoted here by $\bb{R}(\alpha,\beta,\gamma)$. Recall, from \eqref{eq:m3}, that $\bb{R}$ is a block-diagonal matrix. The blocks on the diagonal are square matrices denoted by $\bb{D}_n(\alpha,\beta,\gamma)\in\mathbb{C}^{(2n+1)\times(2n+1)},\,n=0,1,...,N$. A general element in row $m_1$ and column $m_2$ of $\bb{D}_n(\alpha,\beta,\gamma)$ is given by
\begin{align}
	\nonumber\left[\bb{D}_n(\alpha,\beta,\gamma)\right]_{m1}^{m2}&=D_{m_1,m_2}^{n}(\alpha,\beta,\gamma)\\
	&=e^{-jm_1\alpha}d_{m_1,m_2}^n(\beta)e^{-jm_2\gamma},
	\label{eq:a1}
\end{align}
where $d_{m_1,m_2}^n(\beta)$ is the Wigner-d function, which is real-valued and is given by \cite{Varshalovich1988}
\begin{align}
	\nonumber d_{m_1,m_2}^n(\beta)&=\eta_{m_1,m_2}\sqrt{\frac{s!(s+\mu+\nu)!}{(s+\mu)!(s+\nu)!}}\times\\
	&\sin^{\mu}(\beta/2)\cos^{\nu}(\beta/2)P_s^{\mu,\nu}
	(\cos\beta),
\label{eq:a2}
\end{align}
where $\mu=|m_1-m_2|$, $\nu=|m_1+m_2|$, $s=n+(\mu+\nu)/2$, $P_s^{\mu,\nu}(\cdot)$ denotes the Jacobi polynomial, and coefficient $\eta_{m_1,m_2}$ is given by
\begin{equation}
	\eta_{m_1,m_2}=
	\left\{
	\begin{tabular}{cc}
		$1$, & $m_2\geq m_1$\\
		$(-1)^{m_2-m_1}$, & $m_2<m_1$
	\end{tabular}
	\right..
\label{eq:a3}
\end{equation}
Note that the inverse, $\bb{R}^{-1}(\alpha,\beta,\gamma)$, represents a reciprocal rotation and is, indeed, given by a rotation in the reciprocal direction $(-\alpha,-\beta,-\gamma)$:  
\begin{equation}
	\bb{R}^{-1}(\alpha,\beta,\gamma)=\bb{R}(-\alpha,-\beta,-\gamma).
	\label{eq:a4}
\end{equation}
Hence, by substituting $(-\alpha,-\beta,-\gamma)$ into \eqref{eq:a1}, we obtain
\begin{equation}
  \left[\bb{D}_n(-\alpha,-\beta,-\gamma)\right]_{m1}^{m2}=e^{jm_1\alpha}d_{m_1,m_2}^n(-\beta)e^{jm_2\gamma}.
	\label{eq:a5}
\end{equation}
Using \eqref{eq:a4}, the term $d_{m_1,m_2}^n(-\beta)$ can be expressed as
\begin{equation}
		d_{m_1,m_2}^n(-\beta)=(-1)^{|m_1-m_2|}d_{m_1,m_2}^n(\beta).
	\label{eq:a6}
\end{equation}
Next, note that $(-1)^{|m_1-m_2|}\eta_{m_1,m_2}=\eta_{m_2,m_1}$. Now, observe that $\eta_{m_1,m_2}$ is the only non symmetric term with respect to $m_1$ and $m_2$ in \eqref{eq:a2}. This implies that 
\begin{equation}
	d_{m_1,m_2}^n(-\beta)=d_{m_2,m_1}^n(\beta),
\label{eq:a7}
\end{equation}
which, by substituting into \eqref{eq:a5} and recalling that Wigner-d is real valued, yields
\begin{equation}
	\left[\bb{D}_n(-\alpha,-\beta,-\gamma)\right]_{m1}^{m2}=\left(\left[\bb{D}_n(\alpha,\beta,\gamma)\right]_{m2}^{m1}\right)^*.
\label{eq:a8}
\end{equation}
The result in \eqref{eq:a8} implies that
\begin{equation}
	\bb{D}_n(-\alpha,-\beta,-\gamma)=\bb{D}_n^H(\alpha,\beta,\gamma),
	\label{eq:a9}
\end{equation}
which completes the proof, implying that $\bb{R}^{-1}(\alpha,\beta,\gamma)=\bb{R}^H(\alpha,\beta,\gamma)$.
\IEEEQED

	\bibliographystyle{IEEEtran}
	\bibliography{IEEEabrv,taslp2}
\begin{IEEEbiography} 
[{\includegraphics[width=1in,height=1.25in,clip,keepaspectratio]{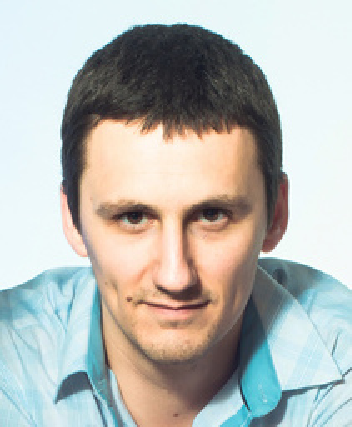}}]{Vladimir Tourbabin}
(S'12) received the B.Sc. degree (summa cum laude) in materials science and engineering and the M.Sc. degree (cum laude) in electrical and computer engineering from Ben-Gurion University of the Negev, Israel, in 2005 and 2011, respectively. He is currently working towards the Ph.D. degree in electrical and computer engineering at Ben-Gurion University.
His current research focuses on audition of humanoid robots.

Mr. Tourbabin is a recipient of the Negev Faran Fellowship.
\end{IEEEbiography}

\begin{IEEEbiography} 
[{\includegraphics[width=1in,height=1.25in,clip,keepaspectratio]{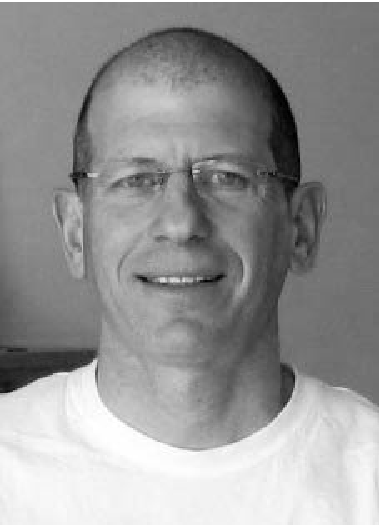}}]{Boaz Rafaely}
Boaz Rafaely (SM’01) received the B.Sc. degree (cum laude) in electrical engineering from Ben-Gurion University, Beer-Sheva, Israel, in 1986; the M.Sc. degree in biomedical engineering from Tel-Aviv University, Israel, in 1994; and the Ph.D. degree from the Institute of Sound and Vibration Research (ISVR), Southampton University, U.K., in 1997. 
At the ISVR, he was appointed Lecturer in 1997 and Senior Lecturer in 2001, working on active control of sound and acoustic signal processing. In 2002, he spent six months as a Visiting Scientist at the Sensory Communication Group, Research Laboratory of Electronics, Massachusetts Institute of Technology (MIT), Cambridge, investigating speech enhancement for hearing aids. He then joined the Department of Electrical and Computer Engineering at Ben-Gurion University as a Senior Lecturer in 2003, and appointed Associate Professor in 2010, and Professor in 2013. 

He is currently heading the acoustics laboratory, investigating sound fields by microphone and loudspeaker arrays. During 2010-2014 he is served as an associate editor for IEEE Transactions on Audio, Speech and Language Processing, and since 2013 as a member of the IEEE Audio and Acoustic Signal Processing Technical Committee. He is currently serving as an associate editor for IEEE Signal Processing Letter and for Acta Acustica united with Acustica, and as a chair of the Israeli Acoustical Association.

Prof. Rafaely was awarded the British Council’s Clore Foundation Scholarship.
\end{IEEEbiography}

\end{document}